\documentclass[]{article}  % for review and submission
\usepackage{graphicx}  % needed for figures
\usepackage{dcolumn}   % needed for some tables
\usepackage{bm}        % for math
\usepackage{amssymb}   % for math

\usepackage{times}
\usepackage{color}

\def\Ld{L_{\rm d}}
\def\Np{N_{\rm p}}

\begin{document}

\begin{center}

{\bf Observation of the Dyakonov--Tamm Wave}\\

Drew Patrick Pulsifer, Muhammad Faryad, Akhlesh Lakhtakia\footnote{To whom correspondence should be addressed; E-mail:  akhlesh@psu.edu}\\

{{\it Department of Engineering Science and Mechanics, Pennsylvania State University,}}\\
{{\it 212 EES Building, University Park, PA 16802, USA}}\\
 \end{center}

{\sf {\bf Abstract.}
A surface electromagnetic wave called the Dyakonov--Tamm wave has been
theoretically predicted to exist at the interface of two dielectric materials at least one of which is both anisotropic and periodically nonhomogeneous. For experimental confirmation, a prism-coupled configuration was used to excite the Dyakonov--Tamm wave guided by the interface of a dense thin film of magnesium fluoride  and a chiral sculptured thin film   of zinc selenide. The excitation was indicated
by a reflection dip (with respect to the angle of incidence in the prism-coupled configuration) that is independent of the polarization state of the incident light as well as the thicknesses of both partnering materials beyond some thresholds. Applications to optical sensing and long-range on-chip communication are expected.
}\\

%\section{\label{sec:level1}First-level heading}
% sections are not used for PRL papers
An electromagnetic surface wave propagates along the interface of two dissimilar materials. The  fields of the surface wave must not only satisfy the Maxwell equations in both partnering materials, but must also decay
far away from the interface~\cite{Boardman,PMLbook,SCbook}. The localization of the fields to the interface makes surface-wave propagation sensitive to changes in the electromagnetic properties of the partnering materials in the region near the interface~\cite{PMLbook,Homolabook}. Such changes  alter the field distribution, the phase speed, and the attenuation rate of the  surface wave and may even cause the surface wave to disappear entirely. This sensitivity is most often exploited for optical sensing applications~\cite{Homolabook,SPL2013,SR2005,KA2007}, but  surface waves also show potential for applications in microscopy, photovoltaics, and communication~\cite{Yeatman,Atwater,Sekhon}.

Most of these applications have been realized for surface-plasmon-polariton (SPP) waves,
which requires one of the two partnering materials to be a metal whereas the other one is a dielectric material.  The concept
of these surface waves excited at 
optical frequencies arose in $1957$, when  Ritchie~\cite{Ritchie} presented a plasma-oscillation explanation for energy losses of fast electrons traversing thin metal films. Simple optical methods to  launch these surface waves emerged shortly thereafter\cite{Turbadar1959,Kretschmann1968,Otto1968}.

Metals dissipate optical energy; hence, most SPP waves do not propagate long distances along the interface plane~\cite{Berini}. The replacement of the partnering metal by a  dielectric  material different from the
other partner should reduce attenuation rates and enhance  propagation distances. Indeed, a second type of electromagnetic surface
wave was predicted in $1977$  to be guided by the interface of two isotropic dielectric materials, at least one of which 
was periodically non-homogeneous in the direction perpendicular to the interface~\cite{YYH}. This     surface wave is often called a Tamm wave as it is analogous to the electronic states predicted to exist at the interface of two crystals by Tamm in $1932$~\cite{Tamm1932}. The experimental observation of Tamm waves followed in 1978~\cite{YYC1978} and, more recently, their application to sensing has been demonstrated as well~\cite{SR2005,KA2007,KA2013}.  A third type of electromagnetic surface wave  was predicted by Dyakonov in $1988$ to be guided by the interface of two homogeneous dielectric materials, at least one of which is anisotropic~\cite{Dyakonov88}. Experimental verification of the
existence of Dyakonov waves came only in 2009, when Takayama \textit{et al.} were able to excite a Dyakonov wave guided by the interface of a liquid and a biaxial dielectric crystal~\cite{Takayama09}. 

Observation of the Dyakonov wave was particularly difficult as it propagates in very narrow ranges of directions in the interface plane. Rarely do the angular sectors of Dyakonon--wave propagation together exceed  $1$~deg of the $360$~deg  available in the interface plane, in currently practical situations~\cite{Takayama08}. Therefore, in 2007 Lakhtakia and Polo~\cite{LP2007} proposed an electromagnetic
 surface wave that is guided by the interface of two dielectric materials, one of which is isotropic and homogeneous and the other is both anisotropic and periodically non-homogeneous in the direction perpendicular to the interface plane.  Combining the attributes of both Tamm and Dyakonov waves, this surface wave was named a Dyakonov--Tamm wave. The angular sectors of Dyakonov--Tamm-wave propagation are so large as to often cover the entire $360$~deg available \cite{PMLbook}.

As the directions of Dyakonov--Tamm-wave propagation are not narrowly restricted, this phenomenon is attractive for both optical sensing and long-range on-chip communication. But, first the Dyakonov--Tamm wave must be observed experimentally. Of the many different combinations
of partnering dielectric materials that will support Dyakonov--Tamm-wave propagation \cite{PMLbook}, that of a homogeneous isotropic
material and a chiral sculptured thin film (STF) is perhaps the most convenient for the experimentalist~\cite{PulsiferJOSAB2013}.

A chiral STF is an assembly of upright and parallel nano-helices~\cite{LMbook}. The nanomaterial is fabricated in a low-pressure chamber containing, most importantly, a source boat, a rotatable platform with a flat face, and a quartz crystal monitor. Material contained in the source boat is evaporated. A collimated portion of the vapor flux is incident on a planar substrate affixed to the platform which is rotated steadily about an axis passing normally through it. The quartz crystal monitor is used to measure the  rate of deposition of the vapor molecules and molecular clusters as a film on the substrate. By controlling the rate at which the vapor condenses onto the substrate, the direction of the collimated vapor flux with respect to the substrate plane, and the rotation speed of the substrate, the structural and optical properties of the chiral STF can be tailored~\cite{LMbook}. The ability to finely and easily tune the structural and optical properties of a chiral STF makes it particularly well suited as a partnering material for launching a Dyakonov--Tamm wave~\cite{PulsiferJOSAB2013}.

\begin{figure}[!ht]
\begin{center}
\includegraphics[width=8.5cm]{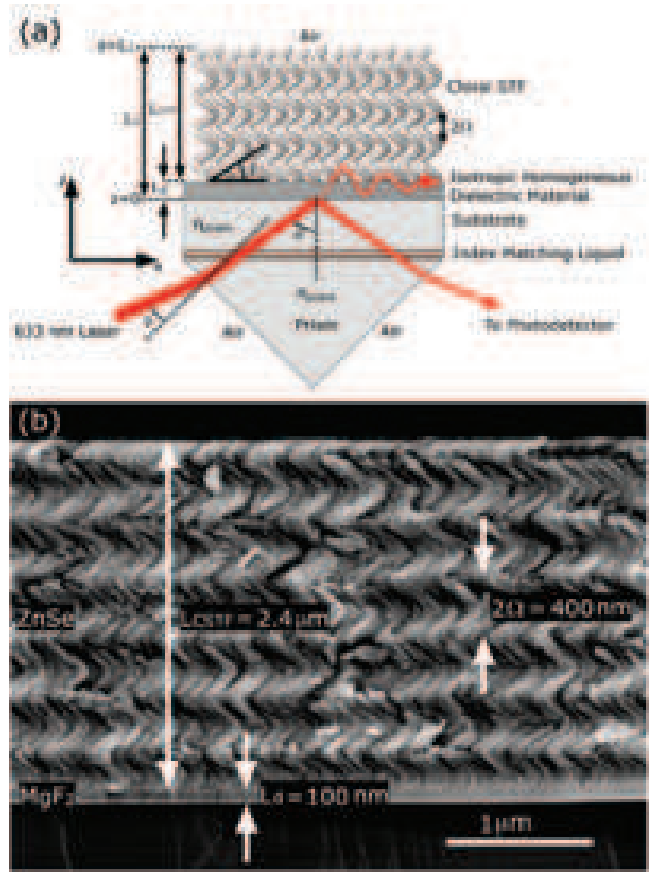}
\caption{(a)~Schematic representation of the  prism-coupled configuration used to excite a Dyakonov--Tamm wave at the interface of a homogeneous isotropic dielectric layer and a chiral STF.  The portion of the path
of light relevant to the identification of surface waves in the prism-coupled configuration
is also shown.   (b)~Cross-sectional field-emission SEM image of the MgF$_2$/ZnSe structure fabricated on a silicon wafer. }
\label{fig1}
\end{center}
\end{figure}

Here we report the first experimental observation of  the Dyakanov--Tamm wave. A prism-coupled configuration was adapted as the method to direct light towards the interface of a chiral STF and a homogeneous, isotropic dielectric material. In this configuration, shown schematically in Fig.~\ref{fig1}a, the hypotenuse of a right-angled isosceles prism is affixed to a substrate by an index-matching fluid. The prism and the substrate are optically matched, both having a refractive index $n_{\rm prism}$. On the other face of the substrate,
a homogeneous, isotropic dielectric material of thickness $\Ld$ and a chiral STF of thickness $L_{\rm CSTF}$ had been deposited earlier. The ratio $\Np=L_{\rm CSTF}/2\Omega$ is an integer~\cite{Np}, where $2\Omega$ is the structural period; then, $N_p$ is the  number of structural periods of the chiral STF.
With the interface of the substrate and the isotropic partnering material serving as the plane $z=0$, the region defined by $z>L_\Sigma=\Ld+L_{\rm CSTF}$ in Fig.~\ref{fig1}a is occupied by air.

%\section*{Fabrication of Samples}
Samples were fabricated by first thermally evaporating an isotropic and homogeneous layer of  magnesium fluoride  onto a dense flint glass (SF11) substrate (Swiss Jewel Co., Philadelphia,
PA, USA) in a custom-made low-pressure chamber. The intended direction of Dyakonov--Tamm-wave propagation was marked by a straight line (the $x$ axis in  Fig.~\ref{fig1}a) on one face of the substrate, and then that face was affixed to the platform face.
 Then the substrate was shuttered, the chamber was evacuated to a pressure of $10$~$\mu$Torr, and a current that slowly increased to $110$~A was passed through a tungsten source boat containing MgF$_2$. The shutter was then removed, and the collimated portion of the MgF$_2$-vapor flux was directed normally towards the substrate which was being rotated at $120$~rpm. The current was adjusted manually to maintain a deposition rate of $\sim 0.4$~nm s$^{-1}$, as indicated by the quartz crystal  monitor, while the layer of thickness $\Ld$
 was being deposited.  After being left in vacuum for several hours in order to cool, the sample was exposed to atmosphere. Then the substrate was rotated about its surface normal by an offset angle $45$~deg, as indicated by parametric simulations~\cite{PulsiferJOSAB2013} to offer the most favorable conditions for the observation of a Dyakanov--Tamm wave.
 The source boat was then  filled with ZnSe, the MgF$_2$-coated substrate was shuttered,  the chamber was evacuated to a base pressure of $0.4$~$\mu$Torr, and the current passing through the source boat was increased slowly to $98$~A.  The substrate was then reoriented so the collimated portion of the ZnSe-vapor flux would be directed at $20$~deg with respect to the substrate plane, a substrate rotation sequence was  initiated, and the shutter was removed. The rotation sequence for the deposition of the chiral STF involved $20$ discrete steps per revolution while holding the substrate stationary for an interval of $75$~s between each movement. As the current manually was adjusted to maintain
 a deposition rate of $0.27$~nm s$^{-1}$,  a structurally right-handed chiral STF with a structural period
$2\Omega\sim400$nm was deposited. Once the desired final thickness of the chiral STF was achieved, the sample was shuttered,  the current passing through the source boat was decreased to $0$~A, and the sample was  left under vacuum for several hours to cool prior to exposing it to atmosphere. Samples of six different types were made:  the magnesium-fluoride layer was either $\Ld\sim100$ or $150$-nm thick, and the zinc-selenide chiral STF had either $\Np=4$, or $5$, or $6$ structural periods.

%\section*{Characterization of Film's Morphology}

Along with each flint-glass substrate, a silicon wafer was  placed on the platform. Thus, the MgF$_2$/ZnSe structure was also deposited on the
silicon wafer for the verification of morphology. The silicon wafer was used
to avoid troubles caused by charging of the SF11 glass substrate  when imaging with a scanning electron microscope (SEM). The wafer 
with the MgF$_2$/ZnSe structure was  immersed in liquid nitrogen and fractured with a punch to reveal a cleaved plane  for cross-sectional images to be obtained. Images were obtained for each combination of films on a field-emission SEM (FEI Nova NanoSEM 630, Hillsboro, OR, USA). As an example, the image in Fig.~\ref{fig1}b reveals a $\sim100$~nm MgF$_2$ layer conjoined with a chiral STF with $N_p=6$ structural periods, each of
thickness $2\Omega\sim400$~nm.   

%\section*{Excitation of the Dyakonov--Tamm Wave}

 The prism-coupled configuration of Fig.~\ref{fig1}a was implemented with a right-angled isosceles prism made of SF11 glass (Edmund Optics, Barrington, NJ, USA). The index-matching fluid had a refractive index of $1.777$ at $633$ nm wavelength, as yielded by the Cauchy equation   formulated using data supplied by the manufacturer (Cargille Laboratories, Cedar Grove, NJ, USA).  
The prism/sample combination was then mounted in a commercial device (Metricon 2010/M, Metricon, Pennington, NJ, USA). A $633$-nm wavelength He-Ne laser (CVI Melles-Griot, Albuquerque, NM, USA) was oriented to ensure that the light incident on the prism would be $p$ polarized (i.e., the magnetic field of the incident light would be aligned with the $y$ axis). 

With $p$-polarized light  incident on a slanted face of the prism at an angle $\phi$,  the intensity $R(\phi)$ of the light exiting the other slanted face of the prism was measured. The angle of incidence $\theta$ on the MgF$_2$/ZnSe structure was computed as $\theta=45^\circ+\sin^{-1}(n_{\rm prism}^{-1}\sin\phi)$, with $n_{\rm prism}= n_{SF11}=1.779$. A half-wave plate was then inserted into the beam-path of the incident laser to convert the $p$-polarized light incident on the prism to the $s$-polarized light (i.e., the electric field would be aligned with the $y$ axis), and the measurement  of the intensity of the exiting light  was repeated.

The sample was then removed from the hypotenuse of the prism and the intensity $R_\circ(\phi)$ of the light exiting the prism was again measured as a function of $\phi$ for each of the two incident polarization states. 
Measured values of the ratio $R(\theta)/R_\circ(\theta)$ were plotted against $\theta$ for all six samples and for both polarization states.
This normalization is meaningful, because total internal reflection occurs at the interface of the prism and air for $\theta\geq\sin^{-1}(1/n_{\rm prism})=34.2$~deg.

%\section*{Results and Discussion}

A sharp dip in the $R(\theta)/R_\circ(\theta)$ vs. $\theta$ curve is indicative of the excitation of a surface wave, provided its location on the $\theta$-axis
is independent of both $\Ld$ and $\Np$ beyond certain threshold values of both parameters \cite{PMLbook,PulsiferJOSAB2013}.
 This is reasonable  because a surface wave is localized to a specific interface, and increasing the thickness of either partnering material
 beyond a threshold will not affect the spatial profiles of the fields of a surface wave. 
 
\begin{figure}[!ht]
\begin{center}
\includegraphics[width=8.5cm]{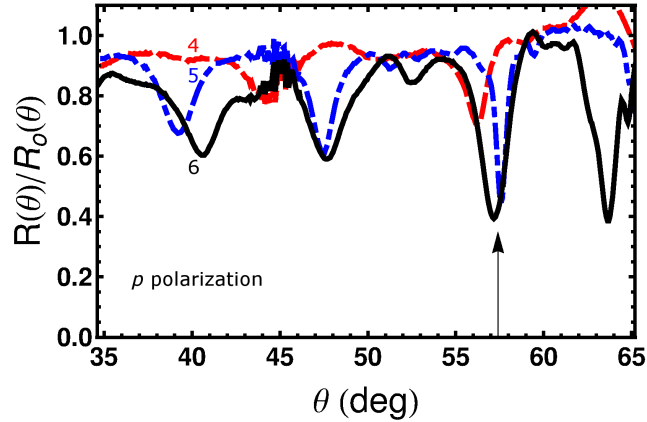}
\caption{Measured values of $R(\theta)/R_\circ(\theta)$ vs. the incidence angle $\theta$ for $\Np= 4$ (red dashed curve), 5 (blue dotted-and-dashed curve), and 6 (black curve), when $\Ld\sim100$~nm and the incident light is $p$ polarized. The excitation of a Dyakonov--Tamm wave
at $\theta\sim57.5$~deg is indicated by the vertical arrow. 
} \label{fig2}
\end{center}
\end{figure}

\begin{figure}[!ht]
\begin{center}
\includegraphics[width=8.5cm]{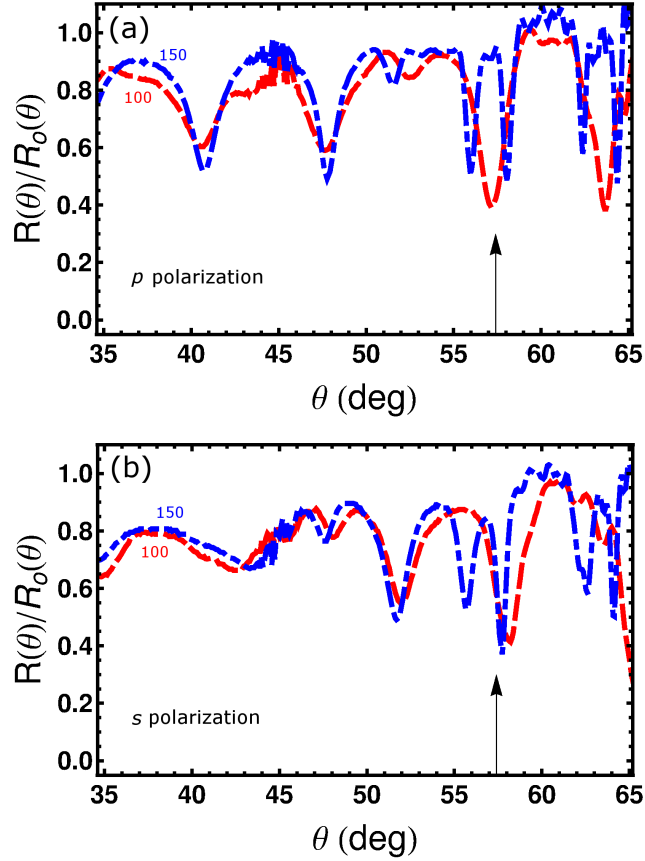}
\caption{Measured values of $R(\theta)/R_\circ(\theta)$ vs. the incidence angle $\theta$ for $\Ld\sim 100$~nm (red dashed curve) and  $150$~nm (blue dotted-and-dashed curve),  when $\Np=6$. The incident light is either (a)  $p$ polarized or (b)  $s$ polarized.  The excitation of a Dyakonov--Tamm wave is indicated by the vertical arrows at $\theta\sim57.5$~deg.
}
 \label{fig3}
 \end{center}
\end{figure}
 
Figure~\ref{fig2} shows $R(\theta)/R_\circ(\theta)$ vs. $\theta$ curves for $\Np\in\left\{4,5,6\right\}$ when $\Ld\sim100$~nm. These three curves have a sharp dip
in $R/R_\circ$ at $\theta\sim57.5$~deg. Aligned dips at $\theta\sim57.5$~deg also occur in Figs.~\ref{fig3}a and b for $\Np=6$ when $\Ld\in\left\{100,150\right\}$~nm for $p$- and $s$-polarization states, respectively. All of these dips indicate that a surface wave is indeed excited at $\theta\sim57.5$~deg.

The question of which interface is supporting the propagation of the surface wave arises. In the prism-coupled configuration, there are three relevant interfaces. The first interface, between the SF11 substrate and the MgF$_2$ layer, can not support the propagation of a surface wave as there exists no solutions to the dispersion equation for that interface \cite[App.~C]{PMLbook}. The other two relevant interfaces are the MgF$_2$/ZnSe interface $z=\Ld$ and the ZnSe/air interface $z=L_\Sigma$. A surface wave guided by either of these two interfaces has to be
a Dyakonov--Tamm wave, by definition. Thus, Figs.~\ref{fig2} and \ref{fig3} are proof of the first observation of the Dyakonov--Tamm wave.

\begin{figure}[!ht]
\begin{center}
\includegraphics[width=8.5cm]{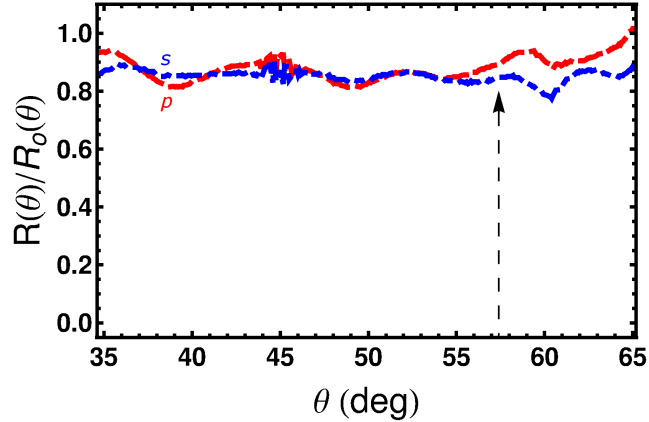}
\caption{Measured values of $R(\theta)/R_\circ(\theta)$ vs. the incidence angle $\theta$ for $p$ polarized (red dashed curve) and  $s$ polarized (blue dotted-and-dashed curve) incident light, when the MgF$_2$ layer is absent and $\Np=5$. There is no dip  at $\theta\sim57.5$~deg, indicating that a Dyakonov--Tamm wave is not excited at the ZnSe/air interface. 
} \label{fig4}
\end{center}
\end{figure}

In order to resolve whether the Dyakonov--Tamm wave in Figs.~\ref{fig2} and \ref{fig3} is guided by the MgF$_2$/ZnSe interface or the
ZnSe/air interface, another experiment was conducted. The sample made for this experiment had $\Np=5$ structural periods of the ZnSe chiral STF, but did not have the MgF$_2$ layer. Because light in the prism is not evanescent when $\theta$ is in a quite large neighborhood of $57.5$~deg, the SF11/ZnSe interface can not guide a Dyakonov--Tamm wave in the prism-coupled experiment conducted with the 
MgF$_2$-deficient sample. Therefore, any sharp dip at $\theta\sim57.5$~deg in the $R(\theta)/R_\circ(\theta)$ vs. $\theta$ curve for both polarization states can only be attributed to a Dyakonov--Tamm wave guided by the ZnSe/air interface. Figure~\ref{fig4} shows the results of this experiment which demonstrates, through the absence of a dip near $\theta\sim57.5$~deg, that the ZnSe/air interface did not guide the Dyakonov--Tamm wave observed in the earlier experiments. We concluded that the Dyakonov--Tamm wave observed in Figs.~\ref{fig2} and \ref{fig3} was guided by the interface of the MgF$_2$ layer (a homogeneous, isotropic dielectric material) and the ZnSe chiral STF (a periodically nonhomogeneous, anisotropic
dielectric material).

%\section*{Conclusions}
The experimental observation of the Dyakonov--Tamm wave, confirming theoretical predictions,
opens a new avenue in the realm of electromagnetic surface waves. Optical sensing and long-range on-chip communication are
among the simpler applications of this new type of surface wave. In particular, as a chiral STF is a porous material that can be
infitrated by solutions of analytes, application for optical
sensing should follow shortly. High sensitivity is expected because the Dyakonov--Tamm wave is expected to propagate
over long distances due to the absence of non-dissipative partnering materials (such as metals) and  thus have a large interaction volume.
Furthermore, as multiple modes of Dyakonov--Tamm wave propagation are possible at the same frequency \cite{MLoc2013},
multi-analyte sensing may be possible, with proper selection of the partnering materials.   

%{\it Acknowledgments}~This work was supported in part by the
%Charles Godfrey Binder Endowment at Penn State.  

\end{document}